\documentstyle[twocolumn,epsf,epsbox]{jpsj}

\def\runtitle{Two-Dimensional $\sigma$-Hole Systems in Boron Layers:
A First-Principles Study on Mg$_{1-x}$Na$_x$B$_2$ and
Mg$_{1-x}$Al$_x$B$_2$}
\def\runauthor{Shugo {\sc Suzuki}, Shin'ichi {\sc Higai}, and Kenji {\sc
Nakao}}

\title
{
\begin{flushleft}
{\small submitted to J. Phys. Soc. Jpn. on 7 Feb 2001}
\end{flushleft}
\vspace*{3mm}
Two-Dimensional $\sigma$-Hole Systems in Boron Layers:\\
A First-Principles Study on Mg$_{1-x}$Na$_x$B$_2$ and
Mg$_{1-x}$Al$_x$B$_2$
}

\author
{ 
Shugo {\sc Suzuki}, Shin'ichi {\sc Higai}$^1$, and Kenji {\sc Nakao}
}

\inst
{
Institute of Materials Science, University of Tsukuba, Tsukuba 305-8573\\
$^1$ National Research Institute for Metals, 1-2-1 Sengen, Tsukuba 305-0047
}

\recdate
{
\today
}

\abst
{ 
We study two-dimensional $\sigma$-hole systems in boron layers by
calculating the electronic structures of Mg$_{1-x}$Na$_x$B$_2$ and
Mg$_{1-x}$Al$_x$B$_2$.  In Mg$_{1-x}$Na$_x$B$_2$, it is found that the
concentration of $\sigma$ holes is approximately described by $( 0.8 +
0.8 x ) \times 10^{22}$ cm$^{-3}$ and the largest attainable
concentration is about $1.6 \times 10^{22}$ cm$^{-3}$ in NaB$_2$.  In
Mg$_{1-x}$Al$_x$B$_2$, on the other hand, it is found that the
concentration of $\sigma$ holes is approximately described by $( 0.8 -
1.4 x ) \times 10^{22}$ cm$^{-3}$ and $\sigma$ holes are disappeared at
$x$ of about 0.6.  These relations can be used for experimental studies
on the $\sigma$-hole systems in these materials.
}

\kword
{
MgB$_2$, NaB$_2$, AlB$_2$, $\sigma$ holes, two dimension, hole
concentration
}

\begin{document}
\sloppy
\maketitle

Quite recently, Nagamatsu et al. have discovered that magnesium
diboride, MgB$_2$, is a superconductor with a high transition
temperature, $T_{\rm c}$, of 39 K \cite{akimitsu01}.  Extensive studies
have now started both experimentally and theoretically.  In particular,
since MgB$_2$ can be regarded as a starting material of undiscovered
high $T_{\rm c}$ superconductors, it is important to search a variety of
materials derived from MgB$_2$.

The structure of MgB$_2$ consists of the layers of triangular lattices
of Mg atoms and the layers of honeycomb lattices of B atoms
\cite{Russell53}.  This is basically the same as that of the
alkali-metal binary graphite intercalation compounds (GIC)
\cite{dresselhaus81}.  Since Mg and B are light elements, where the $s$
and $p$ atomic orbitals play dominant roles, the electronic structure of
MgB$_2$ \cite{Armstrong79} is also very similar to those of GIC
\cite{holzwarth92}.  In spite of these similarity, there are no GIC
superconductors with such a high $T_{\rm c}$; the highest $T_{\rm c}$ of
alkali-metal binary GIC is only about 0.15 K for C$_8$K \cite{tanuma92}.

One of the outstanding differences between MgB$_2$ and GIC is the
existence of the $\sigma$ holes at the center of the Brillouin zone
\cite{Armstrong79}, which are derived from the 2$p_x$ and 2$p_y$ atomic
orbitals of B.  Since the $\sigma$ bands in graphite layers are
energetically very deep, the generation of $\sigma$ holes is extremely
difficult in GIC.  Furthermore, it is interesting to note that holes in
boron layers will show characteristics of two-dimensional (2D) systems.
As revealed so far, 2D systems can provide a rich variety
of physics and possibilities of applications.  It is thus important for
understanding the properties of MgB$_2$ and its derivatives to study the
electronic structures of the $\sigma$-hole systems in boron layers.

In this Letter, we study the 2D $\sigma$-hole systems in
Mg$_{1-x}$Na$_x$B$_2$ and Mg$_{1-x}$Al$_x$B$_2$ at $x$=0, 1/3, 2/3, and
1 by calculating the electronic structures of these materials based on
the density functional theory.  The main result of the present study is
as follows.  In Mg$_{1-x}$Na$_x$B$_2$, since Na is a monovalent metal,
the concentration of the $\sigma$ holes is increased with increasing
$x$, approximately described by $( 0.8 + 0.8 x ) \times 10^{22} {\rm
cm}^{-3}$.  In Mg$_{1-x}$Al$_x$B$_2$, on the contrary, since Al is a
trivalent metal, the concentration of $\sigma$ holes is decreased with
increasing $x$, approximately described by $( 0.8 - 1.4 x ) \times
10^{22} {\rm cm}^{-3}$.  In the latter case, the $\sigma$ holes are
disappeared at $x$ of about 0.6.  These results can be used for
experimental studies on the $\sigma$-hole systems in these materials.

In the present study, we carry out first-principles calculations based
on the density functional theory with the local density approximation
\cite{hohenberg64,kohn65,perdew81,ceperley80}, considering all
electrons.  To check the reliability of the results, the Kohn-Sham
equations are solved by using both the mixed-basis method and the
linear-combination-of-atomic-orbitals (LCAO) method \cite{suzuki00}.  In
this paper, we show the results obtained by the mixed-basis method
although the same results can also be obtained by the LCAO method.  The
cut-off energy used for plane waves is 50 eV and the atomic orbitals
employed as localized orbitals are given in Table I.  We use not only
the atomic orbitals of neutral atoms but also those of charged atoms to
increase the variational flexibility.  The number of used $k$ points in
the full Brillouin zone is 52 for the structure optimization of NaB$_2$
and 185 for the electronic structure calculations of NaB$_2$, MgB$_2$,
and AlB$_2$.  Also, that used in the calculations of
Mg$_{1-x}$Na$_x$B$_2$ and Mg$_{1-x}$Al$_x$B$_2$ is 104.

\begin{halftable}
\caption{Atomic orbitals used for the mixed-basis calculations.}
\label{tab:ao}
\begin{halftabular}{
@{\hspace{\tabcolsep}\extracolsep{\fill}}
cc}\hline
Atom & Atomic orbitals (atomic charge) \\ 
\hline
B  & 1$s$,2$s$,2$p$(neutral) \\
Na & 1$s$,2$s$,2$p$,3$s$(neutral) \\
Mg & 1$s$,2$s$,2$p$,3$s$(neutral); 3$p$(1+); 3$d$(2+) \\
Al & 1$s$,2$s$,2$p$,3$s$,3$p$(neutral); 3$d$(2+) \\
\hline
\end{halftabular}
\end{halftable}

We first calculate the electronic structures of NaB$_2$, MgB$_2$, and
AlB$_2$.  The calculations are performed at the experimental lattice
constants for MgB$_2$ and AlB$_2$; $a$ = 3.084 \AA\ and $c$ = 3.522 \AA\
are used for MgB$_2$ and $a$ = 3.009 \AA\ and $c$ = 3.262 \AA\ are used
for AlB$_2$.  On the other hand, since NaB$_2$ is a hypothetical
material at present, it is necessary to optimize the lattice constants
of this material.  The resultant $a$ and $c$ are 3.02 \AA\ and 4.19 \AA,
respectively, and are used for NaB$_2$.  To check the reliability of
this result, we also optimized the structure of MgB$_2$ and found that
the errors for $a$ and $c$ are $-$2 \% and +0.5 \%, respectively.  We
thus believe that the result for NaB$_2$ is also reliable with the same
accuracy.  The lattice constants $c$ of these materials can be
understood by considering the fact that the ionic radii of Na$^+$,
Mg$^{2+}$, and Al$^{3+}$ are 0.97 \AA, 0.65 \AA, and 0.50 \AA,
respectively.

\begin{figure}
\begin{center}
\epsfigure{file=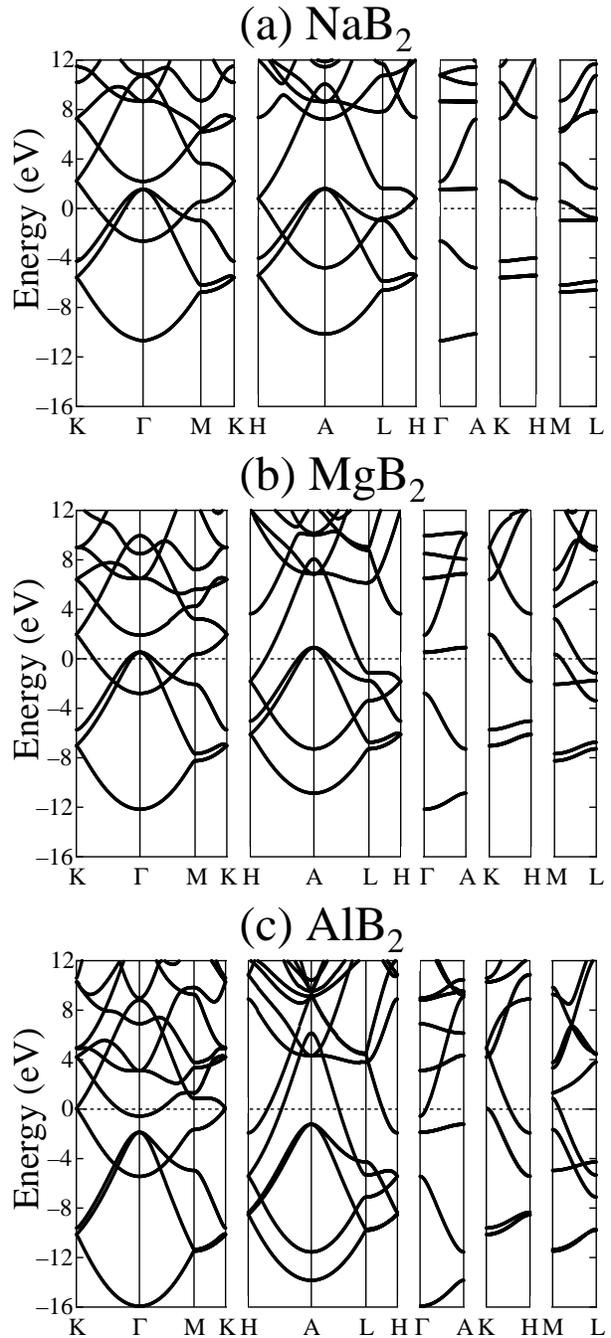,width=8.0cm}
\end{center}
\caption{Band structures of (a) NaB$_2$, (b) MgB$_2$, and (c) AlB$_2$.
 The dotted lines show the Fermi level.} \label{fig:1}
\end{figure}

In Figs. 1(a), 1(b), and 1(c), the calculated electronic structures of
NaB$_2$, MgB$_2$, and AlB$_2$ are shown, respectively.  The dotted lines
in the figures are the Fermi level. The most impressive point is the
position of the top of the $\sigma$ bands derived from the 2$p_x$ and
2$p_y$ atomic orbitals of B.  In NaB$_2$ and MgB$_2$, the top of the
$\sigma$ bands are above the Fermi level and accordingly there exist
$\sigma$ holes in these materials.  This is in strong contrast to the
fact that there are no $\sigma$ holes in GIC.  Since Na is a monovalent
metal while Mg is a divalent metal, the concentration of $\sigma$ holes
is larger in NaB$_2$ than in MgB$_2$.  Thus the concentration of the
$\sigma$ holes can be increased when we increase $x$ in
Mg$_{1-x}$Na$_x$B$_2$.  On the contrary, in AlB$_2$, the top of the
$\sigma$ bands is below the Fermi level and accordingly there are no
$\sigma$ holes in this material.  Thus the $\sigma$ holes are decreased
when we increase $x$ in Mg$_{1-x}$Al$_x$B$_2$ and eventually they are
disappeared at a certain value of $x$.

Furthermore, since the dispersion of the top of the $\sigma$ bands of
all the materials is very small along the $\Gamma$-A direction, the
$\sigma$ holes can show characteristics of 2D systems such as large
fluctuation, etc.  This is in strong contrast to the three
dimensionality of the other carriers in these materials.  In all the
materials, there exist three-dimensional (3D) $\pi$ electrons and/or
holes.  Also, in AlB$_2$, there exist small number of 3D electrons in
the nearly free electron state at the $\Gamma$ point, which is derived
from the hybridization between the 3$s$ atomic orbitals of Al and the
interlayer state of boron layers; this is very similar to the situation
in C$_8$K, where the nearly free electrons also exist at the $\Gamma$
point \cite{Amamiya00}.  It should be noted that, in GIC, there exist
$\pi$ electrons and/or holes and also nearly free electrons and not the
$\sigma$ holes.

\begin{figure}
\begin{center}
\epsfigure{file=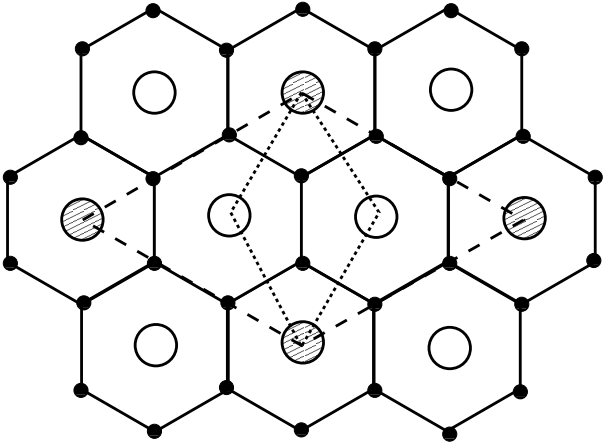,width=8.0cm}
\end{center}
\caption{In-plane $(\sqrt{3} \times \sqrt{3})$ structure (dashed lines)
 used in the electronic structure calculations of Mg$_{1-x}M_x$B$_2$
 ($M$=Na or Al, $x$=0, 1/3, 2/3, and 1) and original $(1 \times 1)$
 structure (dotted lines).  Small closed circles represent B atoms and
 large circles represent the other atoms.  For $x$=0, all the large
 circles are Mg atoms.  For $x$=1/3, open and hatched circles are Mg and
 $M$ atoms, respectively.  For $x$=2/3, open and hatched circles are $M$
 and Mg, respectively.  For $x$=1, all the large circles are $M$ atoms.} 
 \label{fig:2}
\end{figure}

Next, we study the electronic structures of Mg$_{1-x}$Na$_x$B$_2$ and
Mg$_{1-x}$Al$_x$B$_2$ at $x$=0, 1/3, 2/3, and 1.  In the calculations,
we assume the in-plane $(\sqrt{3} \times \sqrt{3})$ structure as shown
in Fig.~2.  This structure has a simplicity that the threefold rotation
axis also exists as in the $(1 \times 1)$ original structure and thus the
same Brillouin zone can be used.  We also assumed that the lattice
constants of these materials can be obtained by linearly interpolating
between the lattice constants of MgB$_2$ and NaB$_2$ or AlB$_2$.  As an
example, we show the result for Mg$_{2/3}$Na$_{1/3}$B$_2$ in Fig.~3.
The obtained electronic structure can be understood by considering the
folding of the original band structures shown in Fig.~1.  The $\sigma$
bands are easily identified by observing that they are the bands with
small dispersion along the $\Gamma$-A direction just above the Fermi
level.

\begin{figure}
\begin{center}
\epsfigure{file=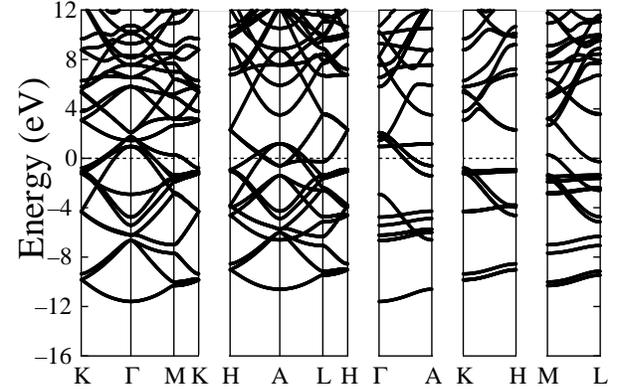,width=8.0cm}
\end{center}
\caption{Band structure of Mg$_{2/3}$Na$_{1/3}$B$_2$. The dotted line
 shows the Fermi level.} \label{fig:3}
\end{figure}

In Fig.~4, we show the top of the $\sigma$ bands in
Mg$_{1-x}$Na$_x$B$_2$ and Mg$_{1-x}$Al$_x$B$_2$ as a function of $x$.
The dotted line in the figure shows the Fermi level.  It is found that
the dependence is monotonous and the largest attainable value is 1.8 eV
for NaB$_2$.  We also find that, in Mg$_{1-x}$Al$_x$B$_2$, the top of
the $\sigma$ bands is below the Fermi level for $x$ larger than about
0.6, that is, the $\sigma$ holes are disappeared for such $x$.  The
dependence on $x$ for the entire region from NaB$_2$ to AlB$_2$ via
MgB$_2$ shown in Fig.~4 cannot be fit with a single straight line.  It
is necessary to fit the result with a curve or, at least, with two
straight lines, one for Mg$_{1-x}$Na$_x$B$_2$ and the other for
Mg$_{1-x}$Al$_x$B$_2$.  If we select the latter choice, the result can
be fit with
\begin{eqnarray}
\label{eq:eq1}
\varepsilon_{\rm top} & = & 0.90 + 0.91 x \; {\rm eV}
\end{eqnarray}
for Mg$_{1-x}$Na$_x$B$_2$ and with
\begin{eqnarray}
\label{eq:eq2}
\varepsilon_{\rm top} & = & 0.90 - 1.57 x \; {\rm eV}
\end{eqnarray}
for Mg$_{1-x}$Al$_x$B$_2$.  Here, we ignore the point for AlB$_2$ in
obtaining the above formula because some quantities, including the top
of the $\sigma$ bands and the cohesive energy as shown below, are not
on the same straight line as the other Mg$_{1-x}$Al$_x$B$_2$ are; this
can be ascribed to the existence of the nearly free electrons in
AlB$_2$ which do not exist in the other Mg$_{1-x}$Al$_x$B$_2$
calculated in the present study.

\begin{figure}
\begin{center}
\epsfigure{file=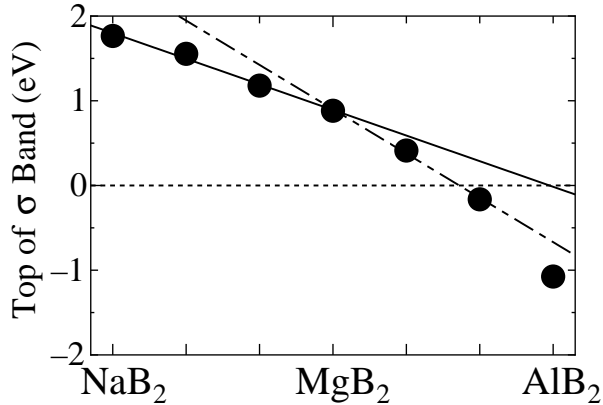,width=8.0cm}
\end{center}
\caption{Top of the $\sigma$ bands as a function of $x$.  The horizontal
 axis is from NaB$_2$ to AlB$_2$ via MgB$_2$.  The dotted line shows the
 Fermi level.  The solid line fits the four points for
 Mg$_{1-x}$Na$_x$B$_2$.  The dotted-dashed line fits the three points
 for Mg$_{1-x}$Al$_x$B$_2$, ignoring the point for AlB$_2$.} 
 \label{fig:4}
\end{figure}

In Fig.~5, we show the cohesive energy of Mg$_{1-x}$Na$_x$B$_2$ and
Mg$_{1-x}$Al$_x$B$_2$ as a function of $x$.  It is found that the most
stable material is AlB$_2$ and the least stable one is NaB$_2$.  The
dependence on $x$ for the entire region from NaB$_2$ to AlB$_2$ via
MgB$_2$ can be described by a single straight line if we ignore the
point for AlB$_2$ because the cohesive energy may be affected by the
existence of the nearly free electrons as mentioned above.  The result
can be fit with
\begin{eqnarray}
\label{eq:eq3}
E_{\rm c} & = & 5.59 - 0.62 x \; {\rm eV/atom}
\end{eqnarray}
for Mg$_{1-x}$Na$_x$B$_2$ and with
\begin{eqnarray}
\label{eq:eq4}
E_{\rm c} & = & 5.59 + 0.62 x \; {\rm eV/atom}
\end{eqnarray}
for Mg$_{1-x}$Al$_x$B$_2$.  Although NaB$_2$ is a hypothetical material
at present, we believe that this material can be synthesized in some
appropriate conditions because the cohesive energy of NaB$_2$, about 5
eV/atom, is not so small; it is almost the same as that of the bulk Si.

\begin{figure}
\begin{center}
\epsfigure{file=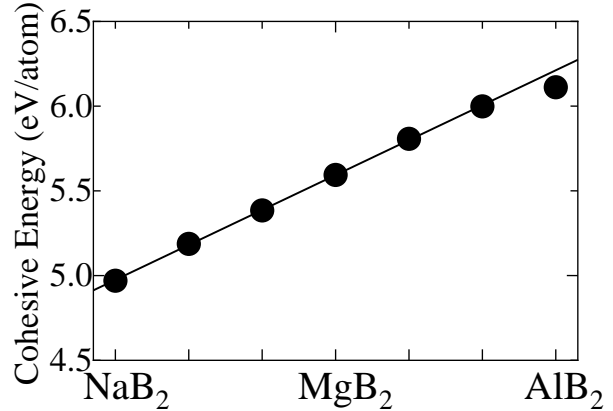,width=8.0cm}
\end{center}
\caption{Cohesive energy as a function of $x$.  The horizontal axis is
 from NaB$_2$ to AlB$_2$ via MgB$_2$.  The solid line fits the six
 points for Mg$_{1-x}$Na$_x$B$_2$ and Mg$_{1-x}$Al$_x$B$_2$, ignoring
 the point for AlB$_2$.} \label{fig:5}
\end{figure}

Next we study the dependence of the concentration of $\sigma$ holes on
$x$, assuming constant density of states for the upper region of the
$\sigma$ bands.  This assumption is good as far as two conditions are
satisfied; one is that the dispersion along the $\Gamma$-A direction is
very small and the other is that the deviation from the in-plane
free-electron-like dispersion is negligible.  Since both conditions are
satisfied as shown in Fig.~1, we derive the formula which give the
concentration of $\sigma$ holes for given $x$.  First, it is necessary
to fit the in-plane free-electron-like dispersion by using effective
masses for heavy and light holes.  As a result, we find that the
effective mass for heavy holes is 0.5$m_{\rm e}$ and that for light
holes is 0.3$m_{\rm e}$, where $m_{\rm e}$ is the mass of free
electrons.  Next, combining with the results shown in Fig.~4, the
following formula are derived for the concentration of the $\sigma$
holes:
\begin{eqnarray}
\label{eq:eq5}
n_{\rm h} & = & ( 0.8 + 0.8 x ) \times 10^{22} \; {\rm cm}^{-3}
\end{eqnarray}
for Mg$_{1-x}$Na$_x$B$_2$ and
\begin{eqnarray}
\label{eq:eq6}
n_{\rm h} & = & ( 0.8 - 1.4 x ) \times 10^{22} \; {\rm cm}^{-3}
\end{eqnarray}
for Mg$_{1-x}$Al$_x$B$_2$.  Thus, in Mg$_{1-x}$Na$_x$B$_2$, the largest
attainable concentration of $\sigma$ holes is about $1.6 \times 10^{22}$
cm$^{-3}$ in NaB$_2$.  In Mg$_{1-x}$Al$_x$B$_2$, on the other hand,
$\sigma$ holes are disappeared at $x$ of about 0.6.

Here, we discuss the possibility of LiB$_2$ as a candidate to increase
the concentration of $\sigma$ holes.  Although one may expect that
LiB$_2$ is the most plausible candidate because of almost the same ionic
radius of Li$^+$, 0.68 \AA, as that of Mg$^{2+}$, this may not the case.
We have found that the structure optimization of LiB$_2$ results in
strong contraction of $c$, which is found to be less than 3 \AA.  The
result strongly conflicts to a simple expectation that the lattice
constant $c$ of LiB$_2$ should be about 3.6 \AA\ if we estimate it by
considering the ionic radius of Li$^+$.  To elucidate stable structure
of LiB$_2$, our study is now in progress.  In spite of this result, one
can still expect that Mg$_{1-x}$Li$_x$B$_2$ for sufficiently small $x$
can be synthesized because the introduction of sufficiently small Li
cannot affect the lattice constant $c$ so strongly.

We next discuss the difference between MgB$_2$ and other metal diborides
such as transition-metal (TM) diborides \cite{Vajeeston01} and
noble-metal diborides, AgB$_2$ and AuB$_2$.  The most important point is
the absence of $d$ atomic orbitals in MgB$_2$ in contrast to the
existence of $d$ atomic orbitals in other metal diborides.  In
particular, since the $d$ atomic orbitals in TM are partly filled, they
form strong covalent bonding with $\sigma$ bonds of boron layers.  This
can destroy 2D $\sigma$-hole system in boron layers; we
have calculated the electronic structures of some TM diborides and have
found that the $\sigma$ bands of boron layers are strongly affected by
the covalent bonding with $d$ atomic orbitals of TM.  On the other hand,
AgB$_2$ and AuB$_2$ can be candidates for similar materials as MgB$_2$.
The reason for this is that the $\sigma$ holes may survive in these
materials because $d$ atomic orbitals in these materials should be
sufficiently lower than the Fermi level and thus the hybridization
between the $d$ atomic orbitals and the $\sigma$ bands may occur at a
lower-energy region.

We finally discuss a possible relation between the superconductivity in
MgB$_2$ and the 2D $\sigma$-hole systems in boron layers.  If the
superconductivity is disappeared in Mg$_{1-x}$Al$_x$B$_2$ at $x$ of
about 0.6, it should be caused by the $\sigma$-hole systems. In
addition, there are no TM diborides with $T_{\rm c}$ as high as that of
MgB$_2$; this may be understood because they have no $\sigma$ holes.  If
this is the case, both of the electron-phonon interaction and the
electron correlation can be the likely origin of the superconductivity.
Since $\sigma$ bonds, especially those of B, C, N, and O, are very
strong, the interaction between the $\sigma$ holes and the in-plane
$\sigma$-bond vibration is expected to be very strong, too.
Furthermore, this electron-phonon coupling can result in a
superconductivity of high $T_{\rm c}$ because the frequency of the
$\sigma$-bond vibration is of the order of 0.2 eV.  On the other hand,
the electron correlation in the $\sigma$ holes seems also important for
the properties of the $\sigma$-hole system in MgB$_2$ because the wave
function of the $\sigma$ holes is localized to considerable degree.

\section*{Acknowledgements}

We would like to thank J. Akimitsu, T. Arima, and H. Tsunetsugu
for useful discussions.

\end{document}